# When should an expert make a prediction?

Yossi Azar[*1], Amir Ban[†1], and Yishay Mansour[‡1,2]

[1]*Blavatnik School of Computer Science, Tel Aviv University*
[2] *Microsoft Research, Hertzelia*

May 23, 2016

**Abstract**

We consider a setting where in a known future time, a certain continuous random variable will be realized. There is a public prediction that gradually converges to its realized value, and an expert that has access to a more accurate prediction. Our goal is to study *when* should the expert reveal his information, assuming that his reward is based on a logarithmic market scoring rule (i.e., his reward is proportional to the gain in log-likelihood of the realized value).

Our contributions are: (1) we characterize the expert's optimal policy and show that it is threshold based. (2) we analyze the expert's asymptotic expected optimal reward and show a tight connection to the Law of the Iterated Logarithm, and (3) we give an efficient dynamic programming algorithm to compute the optimal policy.

## 1 Introduction

Consider a futures market. The traders in a futures market make contracts to buy and sell an asset that will be delivered, and paid for, at a known future date, the *delivery date*. Traders make money by buying for less than the market's spot price on the delivery date, which we shall henceforth call the *true* price or value, or by selling for more. In effect, a futures market is a prediction market for the true price.

Consider now an expert in a futures market. An expert is not a trader himself, but someone who is reputed to have access to a more accurate signal than possessed by regular traders. Often, his reputation and living is based on this. Stock market analysts, investment gurus and various types of journalists fit this description.

The expert contributes to a market by making a public prediction. We assume that the expert's level of expertise, which we measure by *quality* and describe below, is known to the market. Then, such a prediction is a significant market event. Clearly it is optimal for a market to heed an expert whose prediction already encompasses all current common knowledge and adds to it, but this, in itself, does not say of how much value any particular expert announcement was to the market, nor indeed, whether it was positive. It is only at delivery date that the value of an expert's prediction may be evaluated. Market scoring rules, discussed below, show how this may be done in a strategy-proof manner. The expert's reward is proportional to this value. Whether this reward takes the form of actual compensation, or less tangibly in a boost to his reputation as an expert, is immaterial to our discussion.

---
[*]This research was supported in part by The Israeli Centers of Research Excellence (I-CORE) program, (Center No. 4/11), ... , Email: `azar@tau.ac.il`

[†]This research was supported in part by a grant from the Len Blavatnik and the Blavatnik Family Foundation and a grant from the Israel Science Foundation (ISF). Email: `amirban@netvision.net.il`

[‡]This research was supported in part by The Israeli Centers of Research Excellence (I-CORE) program, (Center No. 4/11), by a grant from the Israel Science Foundation (ISF), by a grant from United States-Israel Binational Science Foundation (BSF), and by a grant from the Israeli Ministry of Science (MoS). Email: `mansour.yishay@gmail.com`

While we present our work in the context of a market, the market is not strictly necessary. This work is relevant for any situation where a public is interested in the value of a future continuous variable and has a time-varying consensus estimate of it. Examples abound: The weather or climate, results of sport competitions, election results or new book / movie / album sales.

## 1.1 The Market as a Random Walk

The current price in a futures market represents a current consensus on the true price (assume that interest rates, or inflation rates, have been incorporated into the price). According to the efficient-market hypothesis (EMH), the current price represents all currently available information, and therefore it is impossible to consistently "beat the market". Consistent with the EMH is the random-walk hypothesis, according to which stock market prices (and their derivatives) evolve according to a random walk and thus cannot be predicted. By the random-walk hypothesis, the true price is the result of a random walk from the current market price. Equivalently, and the point of view we take in this paper, the current price is the result of a random walk, reversed in time, from the true price.

A random walk adds periodical (say, daily) i.i.d. steps to the market price. Assuming prices have been adjusted for known trends, the steps have zero mean. By suitable scaling of the price, the step variance can be normalized to 1. Following a common assumption that the random walk is Gaussian, the steps have standard normal distribution (i.e., $N(0, 1)$).

## 1.2 Expert Quality

An expert's expertise consists of having a more accurate signal of the predicted price $x_0$ than the market's, and the expert's quality measures by how much. The quality $q \in [0, 1]$ measures what part of the market's uncertainty the expert "knows", so that it does not figure in the expert's own uncertainty. Equivalently, the expert's uncertainty is $1 - q$ of the market's uncertainty. This proportion is statistical: It is the uncertainties' variances, rather than their realizations, that are related by proportion. If the market price is a Gaussian random walk from the true price with $N(0, 1)$ steps, the expert's prediction is a Gaussian random walk from the true price with $N(0, 1 - q)$ steps.

The expert's knowledge, i.e., the part of the market's uncertainty that the expert is not uncertain about, has steps of zero mean and $q$ variance. On the assumption that the expert's knowledge steps and uncertainty steps are mutually independent, their sum has the sum mean and sum variance of their parts, i.e., they sum back to the market's uncertainty steps of zero mean and variance $q + (1 - q) = 1$.

An expert with $q = 1$ has no uncertainty at all, and his signal equals the true value $x_0$ at all times $t$. At the other extreme, a (so-called) expert with $q = 0$ has no knowledge beyond common knowledge, and his signal equals the market value $x_t$ at all $t$.

In this paper an expert's quality is common knowledge, shared by all traders as well as himself. Whether its value $q$ represents objective reality, or is a belief, based, e.g., on past performance, makes no difference to our discussion.

## 1.3 Scoring a Prediction

A scoring rule is a way to evaluate and reward a prediction of a stochastic event. The predictor declares at time $t > 0$ a probability distribution $p \in \Delta(R)$, and at time 0 some $r \in R$ is realized. A *scoring rule* $S$ rewards the predictor $S(p, r)$ when her prediction was $p$ and the realized value is $r$. In market settings, and many other settings, there exists a current prediction $\bar{p}$ and then the predictor is evaluated on the scoring difference effected $S(p, r) - S(\bar{p}, r)$. Note that the optimization problem of the predictor in a market situation is the same, since he has no influence over $S(\bar{p}, r)$, the only difference is that now the predictor might be penalized for inaccurate predictions. A *proper* scoring rule is a scoring rule for which reporting the true distribution is optimal according to the predictor's information.[1] The logarithmic scoring rule, with $S(p, r) = \log p_r$, scores a prediction by the log-likelihood of the realized value according to the prediction. It is proper, and so is the *Logarithmic Market Scoring Rule* (LMSR) which scores $S(p, r) - S(\bar{p}, r) = \log(p_r/\bar{p}_r)$ when the current prediction is $\bar{p} \in \Delta(R)$. Conditional on $p$ being the correct distribution, the expected reward is the Kullback-Leibler divergence between $p$ and $\bar{p}$: $E_{r \sim p}[\log p_r/\bar{p}_r] = D_{KL}(p||\bar{p})$. While the reward may be positive or negative, its (conditional) expectation is always non-negative.

---

[1]This is subject to the predictor being allowed a single prediction. Chen et al. (2010), for example, show how a predictor's optimal strategy includes bluffing and hiding information when allowed more than one prediction.

In our model expert predictions are scored with LMSR, which the expert maximizes.[2] Chen and Pennock (2010) say "LMSR has become the de facto market maker mechanism for prediction markets. It is used by many companies including Inkling Markets, Consensus Point, Yahoo!, and Microsoft".

## 1.4 The Expert's Dilemma

Assume that the expert has no obligation to speak at any particular time, or at all. He may make a single prediction, at a time of his choosing. What time should the expert choose for his prediction?

The expert is aware at all times of the market current price, $x_t$, and of his own prediction $y_t$, where $t$ is the number of periods remaining to delivery date. He does not know future market prices, or what signals he will have in the future, but he does know that both will converge on the true price, i.e., $x_0 = y_0$. In other words, whatever value he may bring to the market due to his better signal is gradually dissipating, as the market's uncertainty dwindles with the approaching delivery date.

Does that not simply mean that the expert should make a prediction at the earliest opportunity? Not necessarily. What if at time $t$ the expert observes $x_t = y_t$, i.e., that his own prediction coincides with the market's? This may occur by chance even though his signal is, on average, less noisy than the market's. Announcing his prediction will not change the market price, and so its value is minimal. Waiting is the better option, since in all probability, in the next period $(t-1)$, $x_{t-1} \neq y_{t-1}$, and if not then $x_{t-2} \neq y_{t-2}$ etc.. The same may be true if $x_t$ and $y_t$ are not identical but merely close, with what the informal "close" means needing a formal analysis.

## 1.5 Summary of Results

Our results shed an interesting light on the expert's dilemma. Indeed in most cases it is in the expert's advantage to wait for the "right" time, when his prediction and the market's prediction are significantly different. We show that the optimal policy is a time-dependent threshold on the discrepancy between the expert and market signals, i.e., on $|x_t - y_t|$. We also show a near-optimal policy in which the threshold is independent of the current time, depending only on the time horizon $T$ and the expert's quality $q$.

Another conceptual result is the way quality affects the strategy of an expert in revealing his information. High-quality experts will tend to wait more, while lower-quality experts will reveal information earlier. For any time $t$ and discrepancy $|x_t - y_t|$, there is a threshold on the quality $q^*$ such that experts with quality below $q^*$ are advised to reveal their information, while experts with quality above it are advised to wait.

Technically, our analysis of the expert's dilemma shows that the expert strives to maximize a function of the difference between his signal and the market's, namely $(x_t - y_t)^2/t$. The situation he faces is neither a sub-martingale nor a super-martingale, so no easy recipe guides it. We show that his predict now-or-later dilemma is optimally decided by a threshold on the value of $|x_t - y_t|$. This threshold is proportional to a universal function $\theta(t)$ of the time remaining $t$, and to $\sqrt{q}$ (i.e., good experts speak later). His expected reward following this optimal policy is governed by another universal function $\Psi(t)$ and by his quality $q$.

We provide upper and lower bounds for $\Psi(T)$, where $T$ is the total number of time steps, and show that it asymptotically limits at $2 \log \log T$. We provide an efficient dynamic-programming algorithm to compute the $\Psi(t)$ and $\theta(t)$ functions, and also provide a calculation for up to $t = 10^7$, that shows that our asymptotic bounds are fairly tight.

---

[2] We use LMSR to score expert predictions, but not as a market maker mechanism. Since the expert is not a trader, market makers are irrelevant.

## 1.6 Related Work

The Efficient Market Hypothesis was introduced by Fama et al. (1969). The Random Walk Hypothesis is even older, originating in the 19th century, and discussed by, e.g., Samuelson (1965) and Fama (1965), and surveyed in Beechey and Vickery (2000). The Black-Scholes option pricing model Black and Scholes (1973) is based on a Gaussian random walk assumption.

Scoring rules have a very long history, going back to De Finetti (1937), Brier (1950) and Good (1952), and are studied in much subsequent work (Sanders (1963), Winkler (1969), Savage (1971), Gneiting and Raftery (2007)).

Market scoring rules, and the LMSR in particular, were introduced by Hanson (2003) for the study of prediction markets. Much of the literature of prediction markets is concerned with the liquidity of the market, and the need of a market maker to facilitate such liquidity for traders. (See Chen and Pennock (2010) for a survey.) This line of research has unveiled an intriguing connection between the market maker policy and online learning algorithms Chen and Vaughan (2010); Abernethy et al. (2011). Chen et al. (2010) studied the strategy of experts who may predict in a single prediction period, and concluded that, under their model assumptions, experts strive to be the first to make a prediction. Our work uses scoring rules for their original motivation, rather than market maker and liquidity in prediction markets.

Random walks have been thoroughly investigated. We used the textbook Révész (2005) as a general reference. It discusses Khintchine's Law of the Iterated Logarithm Khintchine (1924) at length. However, our optimal policy bound proof is based on a different treatment by Damron (2012).

## 1.7 Paper Organization

The rest of this paper is organized as follows: In Section 2 we describe our model. Section 3 describes the problem an expert faces. In Section 4 we show that the expert's optimal policy is a threshold. Section 5 sets bounds on the optimal policy, and Section 6 shows how to calculate it. In section 7 we summarize and offer concluding remarks.

# 2 Model

## 2.1 Market prediction

A market predicts the outcome of a continuous random variable $X_0$, whose realized value $x_0$ will be revealed at time 0. Time is discrete and flows backwards from an initial period $T$, i.e., $T, \ldots, t, \ldots, 1, 0$. At any time $t > 0$ the market observes $x_0 + \mathcal{Z}_t$ where $\mathcal{Z}_t \sim N(0, t)$. We model $\mathcal{Z}_t$ as a random walk with independent steps $Z_t, \ldots, Z_1$, i.e., $\mathcal{Z}_t = \sum_{\tau=1}^{t} Z_\tau$ and $Z_\tau \sim N(0, 1)$. Let the market prediction (when uninformed by experts) be $X_t := x_0 + \mathcal{Z}_t$ at time $t$, and let $x_t$ be the realized value. With every passing period $t$, the value of $Z_t = z_t$ is revealed and becomes common knowledge, and the market's new prediction changes to $x_{t-1} = x_t - z_t$. Note that the variance of $\mathcal{Z}_t$ decreases with time, and at time 0 the market's prediction coincides with the true value $x_0$. $X_0$ is normally distributed $N(0, \sigma_0^2)$, where we assume $\sigma_0^2 \gg T$. This assumption makes posterior computations dependent solely on observed signals, since[3] we have $E[X_0 | X_t = x_t] = x_t$ and variance $Var(X_0 | X_t = x_t) = t$.

## 2.2 Expert information and goal

There is an expert, with quality $q \in [0, 1]$, whose quality is common knowledge. The expert's quality consists in "knowing" part of the random steps $Z_t$ of every period, and therefore getting a more accurate signal of $X_0$. Formally,

---

[3]When a normal variable with prior distribution $N(0, \sigma_0^2)$ is sampled with known variance $t$ at value $x_t$, its Bayesian posterior distribution is normal with mean $\frac{x_t/t}{1/\sigma_0^2 + 1/t}$ and variance $\frac{1}{1/\sigma_0^2 + 1/t}$. Assuming $\sigma_0^2 \gg T \geq t$, this simplifies to $N(x_t, t)$.

- For every $t$, $Z_t = A_t + B_t$, where $A_t \sim N(0, q)$ and $B_t \sim N(0, 1-q)$ are mutually independent. (Note that $Z_t \sim N(0, 1)$.)

- The expert's private signal at time $t$ is $Y_t = x_0 + B_1 + \ldots + B_t$ and let $y_t$ be its realized value. (Note that if $q = 0$ then $Y_t = X_t$ and if $q = 1$ then $Y_t = x_0$.)

The expert may make a single prediction of the outcome, at a time of his choosing. The expert's predicted distribution at $t$ is $N(y_t, (1-q)t)$. In practice, it is enough for the expert to announce $y_t$ as his entire distribution follows by the model and common knowledge. A prediction's reward is determined at time 0 based on the realized value ($x_0$) by LMSR. (For continuous distributions, the logarithmic scoring rule scores the log of the result probability density). Namely, if the market prediction prior to the expert prediction is $X_t^- \sim N(\mu_-, \sigma_-)$ with density $f_-$, and following the expert prediction the posterior market prediction is $X_t^+ \sim N(\mu_+, \sigma_+)$ with density $f_+$, then the expert reward is $\log(f_+(x_0)/f_-(x_0))$, where $x_0$ is the realized value. An expert who refrains from making a prediction has a benefit of 0. The expert optimization problem is to maximize his expected reward given his private information. *The question before the expert is if and when to make a prediction.*

## 2.3 Preliminaries

The *complementary cumulative distribution function of the standard normal distribution* is conventionally denoted by $\Phi_c(x) := \frac{1}{\sqrt{2\pi}} \int_x^\infty e^{-\frac{t^2}{2}} dt$. The following concentration inequality will be very useful for bounding deviations of a random walk.

**Lemma 2.1.** *Let $\{S_t\}$ be a Gaussian random walk with $N(0, 1)$ steps. For $\lambda \geq 0$*

$$\frac{1}{\lambda + 2} e^{-\frac{\lambda^2}{2}} < \Pr[|S_t| \geq \lambda \sqrt{t}] = 2\, \Phi_c(\lambda) \leq e^{-\frac{\lambda^2}{2}}$$

*Proof.* Formula 7.1.13 from Abramowitz and Stegun (1964) for $x \geq 0$ is

$$\frac{1}{x + \sqrt{x^2 + 2}} < e^{x^2} \int\limits_x^\infty e^{-t^2} dt \leq \frac{1}{x + \sqrt{x^2 + 4/\pi}} \tag{1}$$

Let $z = x\sqrt{2}$, then (1) implies

$$\frac{1}{2(z+2)} < \frac{\sqrt{\frac{2}{\pi}}}{z + \sqrt{z^2 + 4}} < e^{z^2/2} \Phi_c(z) \leq \frac{\sqrt{\frac{2}{\pi}}}{z + \sqrt{z^2 + 8/\pi}} \leq 1/2 \; . \tag{2}$$

As $\Pr[|S_t| \geq \lambda \sqrt{t}] = 2\Phi_c(\lambda)$, the lemma follows. $\square$

# 3 The Expert optimization problem

Consider the case that the expert makes a prediction at time $t$. Let the market prediction prior to the expert prediction be $X_t^- \sim N(\mu_-, \sigma_-^2)$ with density $f_-$ and the posterior market prediction be $X_t^+ \sim N(\mu_+, \sigma_+^2)$ with density $f_+$. Denote expert's reward by $W$.

$$W = \log \frac{f_+(x_0)}{f_-(x_0)} = \log \frac{\frac{1}{\sigma_+ \sqrt{2\pi}} e^{-\frac{(x_0 - \mu_+)^2}{2\sigma_+^2}}}{\frac{1}{\sigma_- \sqrt{2\pi}} e^{-\frac{(x_0 - \mu_-)^2}{2\sigma_-^2}}} = \log \frac{\sigma_-}{\sigma_+} + \frac{(x_0 - \mu_-)^2}{2\sigma_-^2} - \frac{(x_0 - \mu_+)^2}{2\sigma_+^2} \tag{3}$$

As the reward depends on $x_0$, its value is only known at time 0, but the expert can calculate his reward expectation at $t$, based on his belief that $x_0 \sim N(\mu_+, \sigma_+^2)$. This translates to $x_0 - \mu_- \sim N(\mu_+ - \mu_-, \sigma_+^2)$ and $x_0 - \mu_+ \sim N(0, \sigma_+^2)$. As the second moment of the normal distribution $N(\mu, \sigma^2)$ is $\mu^2 + \sigma^2$, we get by taking expectations in (3)

$$\mathop{\mathbb{E}}_{x_0 \sim N(\mu_+, \sigma_+^2)}[W] = \log \frac{\sigma_-}{\sigma_+} + \frac{(\mu_+ - \mu_-)^2 + \sigma_+^2}{2\sigma_-^2} - \frac{0 + \sigma_+^2}{2\sigma_+^2} = \frac{(\mu_+ - \mu_-)^2}{2\sigma_-^2} + \frac{1}{2}\Big(\frac{\sigma_+^2}{\sigma_-^2} - 1 - \log \frac{\sigma_+^2}{\sigma_-^2}\Big)$$

Observe that the right-hand side is the Kullback-Leibler divergence of the two distributions $X_t^+$ and $X_t^-$, i.e.

$$\mathbb{E}[W] = D_{KL}(X_t^+ || X_t^-) = \frac{(\mu_+ - \mu_-)^2}{2\sigma_-^2} + \frac{1}{2}\Big(\frac{\sigma_+^2}{\sigma_-^2} - 1 - \log \frac{\sigma_+^2}{\sigma_-^2}\Big) \tag{4}$$

**Proposition 3.1.** *An expert reward expectation when making a prediction is*

$$\mathbb{E}[W] = \frac{(y_t - x_t)^2}{2t} - \frac{1}{2}\Big(q + \log(1-q)\Big) \tag{5}$$

*Proof.* For this setting, we have $\mu_- = x_t$, $\sigma_-^2 = t$, $\mu_+ = y_t$, and $\sigma_+^2 = (1-q)t$. Substituting these in (4) we derive (5). $\square$

Note that the reward $W$ may be positive or negative depending on $x_0$, but its expectation is always non-negative.

Considering the expert's expected reward (5), we observe that the right-hand side has a term $-\frac{1}{2}\big(q + \log(1-q)\big)$ which depends only on the expert's quality, and is strictly positive for $0 < q < 1$. The other term is non-negative. Therefore, an expert with positive quality will always make a prediction, in the last period at the latest. The expert maximizes his expected reward by selecting a prediction time $t$ that maximizes the other term $\frac{(y_t - x_t)^2}{2t}$.

At any time $t$, the expert knows $x_t$ and $y_t$, and can calculate his expected reward from an immediate prediction at time $t$. However, the expert does not know $y_\tau$ and $x_\tau$ for any $\tau < t$. *Should the expert make his prediction now, or wait for a higher-benefit opportunity?*

Let $S_t = (Y_t - X_t)/\sqrt{q} = (A_1 + \ldots + A_t)/\sqrt{q}$, hence the series $\{S_t = (Y_t - X_t)/\sqrt{q}, t = 1, 2, \ldots\}$ is a Gaussian random walk with $N(0, 1)$ i.i.d. steps. The random variable $S_t^2/t$ is an affine transformation of the expert's reward (5), so the optimal policy to maximize it is essentially identical to the optimal expert prediction policy. From now on we consider maximizing $S_t^2/t$, which we call the *Canonical Problem*. For clarity, the following is an explicit statement of the problem.

**Problem 1** (Canonical Problem). *Let $S_t$ be a Gaussian random walk with $N(0,1)$ steps. Suppose an expert is successively presented with $S_T, S_{T-1}, \ldots, S_1$. If the expert stops at $t$, his reward is $S_t^2/t$. How should the expert maximize his reward?*

We define the expected reward of an expert who follows the optimal strategy in the canonical problem, distinguishing between the expectation given the current value of $S_t = c$, denoted by $\psi_t(c)$, and the expectation independent from the current value, denoted by $\Psi(t) = \mathbb{E}_c[\psi_t(c)]$.

**Definition 1.** *Denote by $\psi_t(c)$ the expert's optimal-strategy expected reward at $t$, conditional on $S_t = c$. The expert's optimal-strategy expected reward $\Psi(t)$, is*

$$\Psi(t) := \mathop{\mathbb{E}}_{c \sim N(0,t)}[\psi_t(c)] = \frac{1}{\sqrt{2\pi t}} \int_{-\infty}^{+\infty} e^{-\frac{c^2}{2t}} \psi_t(c) dc = \frac{1}{\sqrt{2\pi}} \int_{-\infty}^{+\infty} e^{-\frac{x^2}{2}} \psi_t(x\sqrt{t}) dx$$

# 4 The Optimal Policy is a Threshold

In this section we derive properties of the optimal strategy of the expert. In Proposition 2 we give a recursive formula for the canonical problem's expected reward $\psi_t(c)$, and state some properties satisfied by the function.

Later, in Proposition 3, we show that the expert's optimal policy is a threshold policy. Namely, if he stops for some value $c$ he will also stop for any value $c' > c$. We also show that there is always a finite threshold, namely, for any time $t$ there is some value $\theta(t)$ such that for $c > \theta(t)$ the expert stops.

We conclude with Proposition 4 that relates the canonical problem back to the expert's strategy, showing that the expert strategy depends on $(x_t - y_t)^2$. An implication of this is the interesting insight that experts with higher quality will wait with their prediction and pursue higher rewards in situations where their lower-quality peers will not.

**Proposition 4.1.**    *1. The optimal expected reward $\psi_t(\cdot)$ satisfies the recursion formula*

$$\psi_t(c) = \max\left[\frac{c^2}{t}, \psi_t^{WAIT}(c)\right] \tag{6}$$

*where $\psi_t^{WAIT}(c)$ is the expectation from waiting at least one period, and is equal to,*

$$\psi_t^{WAIT}(c) = \frac{1}{\sqrt{2\pi\frac{t-1}{t}}} \int\limits_{-\infty}^{\infty} e^{-\frac{(x-\frac{c}{t})^2}{2\frac{t-1}{t}}} \psi_{t-1}(c-x)dx \tag{7}$$

$$= \frac{1}{\sqrt{2\pi}} \int\limits_{-\infty}^{\infty} e^{-\frac{x^2}{2}} \psi_{t-1}\left(\frac{t-1}{t}c - \sqrt{\frac{t-1}{t}}x\right)dx \tag{8}$$

$$= \frac{1}{\sqrt{2\pi\frac{t-1}{t}}} \int\limits_{-\infty}^{\infty} e^{-\frac{(x-\frac{t-1}{t}c)^2}{2\frac{t-1}{t}}} \psi_{t-1}(x)dx \tag{9}$$

*2. The conditional optimal-strategy expected reward $\psi_t(c)$ is Lipschitz continuous, piecewise differentiable, positive, even and increasing in $|c|$.*

*Proof.* We first prove the first part of the Proposition. The expert's expectation is the maximum between the benefit for stopping (which is $c^2/t$, since $S_t = c$), and the benefit for waiting at least one period. If the expert chooses to wait, and the value of $S_{t-1}$ is $c - x$, his benefit will be $\psi_{t-1}(c-x)$. For every real $x$, the probability density for this is

$$\Pr[S_{t-1} = c - x | S_t = c] = \frac{\Pr[S_{t-1} = c - x]\Pr[S_t - S_{t-1} = x]}{\Pr[S_t = c]}$$

$$= \frac{1}{\sqrt{2\pi\frac{t-1}{t}}} \exp\left\{-\frac{1}{2}\left(\frac{(c-x)^2}{t-1} + x^2 - \frac{c^2}{t}\right)\right\}$$

$$= \frac{1}{\sqrt{2\pi\frac{t-1}{t}}} \exp\left\{-\frac{(x-\frac{c}{t})^2}{2\frac{t-1}{t}}\right\},$$

and so the expectation for waiting is $\psi_{t-1}(c-x)$ averaged over this conditional probability for each $x$,

$$\psi_t^{WAIT}(c) = \frac{1}{\sqrt{2\pi\frac{t-1}{t}}} \int\limits_{-\infty}^{\infty} e^{-\frac{(x-\frac{c}{t})^2}{2\frac{t-1}{t}}} \psi_{t-1}(c-x)dx \tag{10}$$

This completes the proof of (7). (8) is derived from (10) by substituting $x \to \sqrt{\frac{t-1}{t}}x + \frac{c}{t}$, while (9) is derived from it by substituting $x \to c - x$.

We now prove the second part of the Proposition. We prove the claim by induction on $t$. For $t = 1$, the expert stops for every $c$ so $\psi_1(c) = c^2$ for which the claim holds. Assume the claim holds for $t - 1$. Then by (6) and (7) $\psi_t(c)$ is positive, even, Lipschitz continuous and piecewise differentiable. It remains to show that $\psi_t$ is increasing in $|c|$. In any range where $\psi_t(c) = c^2/t$ it is increasing in $|c|$. Elsewhere, we differentiate (9) with respect to $c$,

$$\frac{d}{dc}\psi_t(c) = \frac{1}{\sqrt{2\pi\frac{t-1}{t}}} \int_{-\infty}^{\infty} (x - \frac{t-1}{t}c)e^{-\frac{(x-\frac{t-1}{t}c)^2}{2\frac{t-1}{t}}}\psi_{t-1}(x)dx$$

$$= \frac{1}{\sqrt{2\pi\frac{t-1}{t}}} \int_{-\infty}^{\infty} xe^{-\frac{x^2}{2\frac{t-1}{t}}}\psi_{t-1}\left(\frac{t-1}{t}c + x\right)dx$$

$$= \frac{1}{\sqrt{2\pi\frac{t-1}{t}}} \int_{0}^{\infty} xe^{-\frac{x^2}{2\frac{t-1}{t}}}\left[\psi_{t-1}\left(\frac{t-1}{t}c + x\right) - \psi_{t-1}\left(\frac{t-1}{t}c - x\right)\right]dx$$

where the second identity uses the substitution $x \to \frac{t-1}{t}c + x$.

For $x \geq 0, c \geq 0$, we have that $|\frac{t-1}{t}c + x| \geq |\frac{t-1}{t}c - x|$. As by the induction hypothesis $\psi_{t-1}(c)$ is non-decreasing in $|c|$, the integrand is non-negative, and so is the integral. Hence, $\psi_t'(c) \geq 0$. □ □

Next we state and prove a proposition that says that the optimal policy is a threshold, and show that for every $t$ the threshold is finite.

**Proposition 4.2.** *Define $\theta(t)$ to be the smallest $c \geq 0$ for which $\psi_t(c) = c^2/t$. For every $t$, $\theta(t)$ exists, and the optimal strategy in the canonical problem is to stop at the first $t$ for which $|S_t| \geq \theta(t)$.*

*Proof.* We need to show that there is a $c$ such that $\psi_t(c) = c^2/t$, and that it is optimal to stop for $|S_t| \geq c$. The latter claim we show by showing that if it is optimal to stop for $|S_t| = c$ then it is also optimal for $|S_t| > c$. We prove both parts of this claim by induction on $t$, the first part by proving a stronger statement, that $\psi_t(c) - c^2/t$ is non-increasing in $|c|$.

For $t = 1$, $\theta(1) = 0$, i.e., the expert stops for any value of $S_1$, and $\psi_1(c) - c^2 = 0$ is constant and so non-increasing.

Our induction hypothesis is that $\theta(t - 1)$ exists, and that $\psi_{t-1}(c) - c^2/(t - 1)$ is non-increasing in $|c|$. By Proposition 2 $\psi_{t-1}$ is an even function, hence, its derivative is an odd function. For $c \geq 0$ we have $\psi_{t-1}'(c) - (2c)/(t - 1) \leq 0$ and for $c < 0$, $\psi_{t-1}'(c) - (2c)/(t - 1) = \psi_{t-1}'(-|c|) + (2|c|)/(t - 1) = -\psi_{t-1}'(|c|) + (2|c|)/(t - 1) \geq 0$. Therefore,

$$|\psi_{t-1}'(c)| \leq \frac{2|c|}{t - 1} \tag{11}$$

Within any segment in which $\psi_t(c) = c^2/t$, $|\psi_t'(c)| = \frac{2|c|}{t}$. Elsewhere, i.e., where $\psi_t(c) > c^2/t$, we

differentiate (8) with respect to $c$, then substitute $x \to \sqrt{\frac{t-1}{t}}c - \sqrt{\frac{t}{t-1}}x$

$$\psi'_t(c) = \frac{1}{\sqrt{2\pi}} \frac{t-1}{t} \int\limits_{-\infty}^{\infty} e^{-\frac{x^2}{2}} \psi'_{t-1}\Big(\frac{t-1}{t}c - \sqrt{\frac{t-1}{t}}x\Big) dx$$

$$= \frac{1}{\sqrt{2\pi}} \sqrt{\frac{t-1}{t}} \int\limits_{-\infty}^{\infty} \exp\Big\{ -\frac{(\frac{t-1}{t}c-x)^2}{2\frac{t-1}{t}} \Big\} \psi'_{t-1}(x) dx$$

$$= \frac{1}{\sqrt{2\pi}} \sqrt{\frac{t-1}{t}} \int\limits_{0}^{\infty} \Big[ \exp\Big\{ -\frac{(\frac{t-1}{t}c-x)^2}{2\frac{t-1}{t}} \Big\} - \exp\Big\{ -\frac{(\frac{t-1}{t}c+x)^2}{2\frac{t-1}{t}} \Big\} \Big] \psi'_{t-1}(x) dx \ ,$$

where the last equality relies on the fact that $\psi'_{t-1}$, the derivative of an even function, is odd. As for $c \geq 0, x \geq 0$, $\exp\Big\{ -\frac{(\frac{t-1}{t}c-x)^2}{2\frac{t-1}{t}} \Big\} - \exp\Big\{ -\frac{(\frac{t-1}{t}c+x)^2}{2\frac{t-1}{t}} \Big\} \geq 0$, we can apply (11) to derive (when $\psi_t(c) > c^2/t$)

$$\psi'_t(c) \leq \frac{1}{\sqrt{2\pi}} \sqrt{\frac{t-1}{t}} \int\limits_{0}^{\infty} \Big[ \exp\Big\{ -\frac{(\frac{t-1}{t}c-x)^2}{2\frac{t-1}{t}} \Big\} - \exp\Big\{ -\frac{(\frac{t-1}{t}c+x)^2}{2\frac{t-1}{t}} \Big\} \Big] \frac{2x}{t-1} dx$$

$$= \frac{1}{\sqrt{2\pi}} \sqrt{\frac{t-1}{t}} \int\limits_{-\infty}^{\infty} \exp\Big\{ -\frac{(\frac{t-1}{t}c-x)^2}{2\frac{t-1}{t}} \Big\} \frac{2x}{t-1} dx$$

$$= \frac{t-1}{t} \frac{2\frac{t-1}{t}c}{t-1} = \frac{2(t-1)c}{t^2}$$

In summary, whenever $\psi_t(c) > c^2/t$ we have,

$$|\psi'_t(c)| \leq \frac{2(t-1)|c|}{t^2} < \frac{2|c|}{t} \ . \tag{12}$$

We conclude $|\psi'_t(c)| \leq \frac{2|c|}{t}$ for every $c$ and so $\psi_t(c) - c^2/t$ is non-increasing in $|c|$. Since $\psi_t(c) - c^2/t \geq 0$, and, if $\theta(t)$ exists, $\psi_t(\theta(t)) - \theta^2(t)/t = 0$, we conclude $\psi_t(c) - c^2/t = 0$ for all $|c| \geq \theta(t)$. I.e., the expert should stop whenever $|S_t| \geq \theta(t)$.

It remains to be shown that $\theta(t)$ exists. To show that, it is enough to show that for some $c'$, stopping is at least as beneficial as waiting, i.e., that

$$\psi_t^{WAIT}(c') \leq \frac{c'^2}{t} \ . \tag{13}$$

By (12), for $c \geq 0$

$$\psi_t^{WAIT}(c) \leq \psi_t(0) + \int\limits_{0}^{c} \frac{2(t-1)|x|}{t^2} dx = \psi_t(0) + \frac{(t-1)c^2}{t^2} \ .$$

Substituting any $c' \geq t\sqrt{\psi_t(0)}$ in the above satisfies (13), since we have $\psi_t^{WAIT}(c') \leq t\psi_t(0) \leq c'^2/t$. Therefore,

$$\theta(t) \leq t\sqrt{\psi_t(0)} \ . \tag{14}$$

$$\square \qquad\qquad\qquad\qquad \square$$

The expert's optimal policy and expectation now follows

**Proposition 4.3.** *1. At time $t$, a prediction made by an expert with quality $q$ maximizes his expected reward if and only if*

$$(y_t - x_t)^2 \geq q\theta^2(t)$$

*2. The expert's expected reward from following this policy for every $t$ is*

$$\mathbb{E}[W|y_t, x_t] = \frac{1}{2}q\psi_t\left(\frac{y_t - x_t}{\sqrt{q}}\right) - \frac{1}{2}\left(q + \log(1-q)\right) \tag{15}$$

*where $x_t$ is the market prediction and $y_t$ the expert's prediction at time $t$.*

*Proof.* As noted in Section 3, $(Y_t - X_t)/\sqrt{q}$ is a Gaussian random walk with $N(0,1)$ steps. Using the solution of the Canonical Problem, the expert should predict when $|(y_t - x_t)/\sqrt{q}| > \theta(t)$. In other words, when $(y_t - x_t)^2 > q\theta^2(t)$.

The expert's expected reward was given in (5), where it was noted that the time-dependent term is $(y_t - x_t)^2/(2t)$. In the Canonical Problem, he would be maximizing $(y_t - x_t)^2/(qt)$ for an expected gain of $\psi_t(c) = \psi_t\left(\frac{y_t - x_t}{\sqrt{q}}\right)$. In his actual prediction problem, he would be maximizing $(y_t - x_t)^2/(2t)$, i.e., $q/2$ times his Canonical Problem target. $\qquad\square$

We observe from Proposition 4 that given the same observations $y_t, x_t$, experts with different qualities may make different decisions regarding the timing of predictions

- Experts with $q < \frac{(y_t - x_t)^2}{\theta^2(t)}$ will make a prediction.

- Experts with $q > \frac{(y_t - x_t)^2}{\theta^2(t)}$ will wait.

Thus a high-quality expert will remain silent in a situation where a low-quality expert will speak. However, since qualities are limited to 1, all experts, regardless of their quality, should make a prediction when

$$|y_t - x_t| > \theta(t)$$

# 5 Bounds on the Optimal Policy

In this section we derive upper and lower bounds on the expectation of the reward of the optimal policy. The proof is based on proofs of the *Law of the Iterated Logarithm* (LIL), but with several important differences.

The LIL (Hartman and Wintner (1941)) states that in a random walk, where the increments are $N(0,1)$, the maximum deviation is $\sqrt{2t\log\log t}$ with probability 1. Namely, for any $\epsilon > 0$, the deviation, with probability one, is greater than $\sqrt{2(1-\epsilon)t\log\log t}$ at infinitely many times $t$, and only at finite number of times $t$ greater than $\sqrt{2(1+\epsilon)t\log\log t}$.

In our proof, we consider finite error bounds, and are not satisfied with *probability one* results. For the optimal policy, in our setting, it is sufficient to have a single high value, and there is no need to have the event occur infinitely often. This implies that we need to consider a more refined bound.

Another difference is that we consider the deviations for all $t \leq T$ for a given $T$. When we consider the expert utility, its expected value is with respect to the global time horizon $T$, and not with respect to the current time. Technically, this results in a different threshold from the LIL: We set the threshold to be $\Phi(t) := \sqrt{2t\log\log T}$, while the LIL sets $\sqrt{2t\log\log t}$. The reason for this is that we are interested in bounding an expert's expectation as a function of the time horizon $T$. Proving that the LIL bound is exceeded (or not) does not serve to establish a bound on the expert's expectation at $T$, since the benefit is a function of $t$, and not of $T$. On the other hand, assigning a probability that a deviation of $\Phi(t)$ is exceeded is equivalent to assigning a probability that the expert's reward exceeds $\Phi^2(t)/t = 2\log\log T$, a function of $T$.

Our bound $\Phi(t)$ is *higher* than the LIL's bound of $\sqrt{2t \log \log t}$, as $T \geq t$. Exceeding it does not contradict the LIL, first, since it might exceed only once and not infinitely often, and second, since it is not claimed that for any time $t$ in which it holds we have $\log \log T / \log \log t > 1 + \epsilon$ with probability 1 for some fixed $\epsilon > 0$.

Recall that $\Psi(T)$ is the expected value of the optimal policy for a time horizon $T$. The following proposition places upper and lower bounds on the expectation of the optimal policy, i.e., $\Psi(T)$.

**Proposition 5.1.** *For every $\epsilon > 0$ and $T > 10$*

$$\Psi(T) \leq (1 + \epsilon) 2 \log \log T + \gamma_1(T, \epsilon)$$

*and for every $1/2 > \epsilon > 0$ and $T > 16$*

$$\Psi(T) \geq (1 - \gamma_2(T, \epsilon))(1 - \epsilon) 2 \log \log T$$

*where*

$$\gamma_1(T, \epsilon) := \frac{12}{\log\left(1 + \frac{(\sqrt{1+\epsilon} - \sqrt{1+\epsilon/2})^2}{1+\epsilon/2}\right)} \frac{1}{(\log T)^{\epsilon/2}}$$

$$\gamma_2(T, \epsilon) := \exp\left(-\frac{(\log T)^{\epsilon/4 - \epsilon^2/64}}{(1 - \epsilon/8)\sqrt{2 \log \log T} + 2} \cdot \frac{1}{\log \frac{20}{\epsilon^2}}\right) + \frac{1}{(\log T)^{\epsilon/8}}$$

In order to get a better understanding of the magnitude of the $\gamma$s note that for $\epsilon \in [0, 1]$ we have,

$$\gamma_1(T, \epsilon) < \frac{12}{\log(1 + \epsilon^2/4)} \frac{1}{(\log T)^{\epsilon/2}} < \frac{96}{\epsilon^2} \frac{1}{(\log T)^{\epsilon/2}} = O\left(\frac{1}{\epsilon^2 (\log T)^{\epsilon/2}}\right) \tag{16}$$

and for $\epsilon < 0.2$ and $\log \log T > 16/\epsilon^2$ we have

$$\gamma_2(T, \epsilon) = O(\exp\{-(\log T)^{\epsilon/16}\})$$

and in fact we can make the exponent $\epsilon/16$ much closer to $\epsilon/4$.

The proof of Proposition 5 will be done in two parts. First, for the upper bound, in Lemma 5.1. Second, for the lower bound, in Lemma 5.3. We start with the upper bound.

**Lemma 5.1.** *For every $\epsilon > 0$ and $T > 10$,*

$$\Psi(T) \leq (1 + \epsilon) 2 \log \log T + \gamma_1(T, \epsilon)$$

*where*

$$\gamma_1(T, \epsilon) := \frac{12}{\log\left(1 + \frac{(\sqrt{1+\epsilon} - \sqrt{1+\epsilon/2})^2}{1+\epsilon/2}\right)} \frac{1}{(\log T)^{\epsilon/2}}$$

Here is an overview of the proof, which is found in the Appendix: Instead of bounding $\Psi(T)$, we bound a larger quantity, the expected value of $M_T$, defined as the maximum of $S_t^2/t$ for $t \leq T$. For this we need to bound the probability that $M_T > (1 + \epsilon) 2 \log \log T$, and the expectation of $M_T$ will follow by integrating this probability over $\epsilon$. Therefore our probability bound must also be tight enough to assure that this integral does not diverge.

We partition the time $[1, T]$ to a logarithmic number of gaps, with endpoints $a^k$ where $a = 1 + \Theta(\epsilon^2)$. This implies that we have $\log_a(T) = \Theta(\epsilon^{-2} \log T)$ such endpoints. We show that with high probability, for any endpoint $a^k$ the probability that the deviation is more than $\sqrt{2a^k \log \log T}$ times $1 + \Theta(\epsilon)$ is significantly less than $1/(\log_a T)$. Since there are $\log_a T$ such endpoints, a union bound makes it hold for all endpoints by a probability close to 1. The next step is to bound the deviation within the gaps $(a^k, a^{k+1})$. For this we use an inequality attributed to Levy which relates the probability of a deviation of each single time to the probability of deviation of the maximum over all the time points. Our bound for $\Pr[M_T > (1+\epsilon)2\log\log T]$ is a union bound of the endpoints bound and the gaps bound. This gives us a high probability bound for $M_T$. To complete the proof, we integrate over the failure probabilities to get an upper bound on the expectation of $M_T$.

We establish a lower bound of for $\Psi(T)$ by first lower bounding $M_T$.

**Lemma 5.2.** *For every $0 < \epsilon < 1/2$ and every $T > 16$*

$$\Pr[M_T > 2(1-\epsilon)\log\log T] \geq 1 - \gamma_2(T, \epsilon)$$

The proof of this lemma is in the Appendix. Here is an overview of it: We show that with high probability $M_T$ is large enough. This will establish a lower bound for a simple, non-optimal policy and therefore also a lower bound for $\Psi(T)$.

As in the upper bound, to bound $M_T$ we again partition the time $[1, T]$ to a logarithmic number of gaps, with endpoints $a^k$ where $a > 1$. (Unlike the case of the upper bound, where $a = 1+\Theta(\epsilon)$, for the lower bound we use $a = \Theta(\epsilon^{-2})$, and hence there are huge gaps between $a^k$ and $a^{k+1}$.) We lower-bound the deviation within each gap, and, using the independence of the gaps establish that with probability close to 1 one of the gaps $(a^k, a^{k+1})$ will have a large enough relative deviation. We also show that, with high probability, no endpoint $S_{a^k}$ is "too negative". Combining the large deviation within the gap and its not-too-negative value at its start leads to our desired bound.

We can now complete the lower bound proof for $\Psi(T)$.

**Lemma 5.3.** *For every $1/2 > \epsilon > 0$ and $T > 16$,*

$$\Psi(T) \geq (1 - \gamma_2(T, \epsilon))(1 - \epsilon)2\log\log T$$

*where*

$$\gamma_2(T, \epsilon) := \exp\left(-\frac{(\log T)^{\epsilon/4 - \epsilon^2/64}}{(1 - \epsilon/8)\sqrt{2\log\log T} + 2} \cdot \frac{1}{\log\frac{20}{\epsilon^2}}\right) + \frac{1}{(\log T)^{\epsilon/8}}$$

*Proof.* Lemma 5.2 suggests the following stopping strategy: Choose $0 < \epsilon < 1/2$, and use a stopping strategy of $S_t^2/t > (1-\epsilon)2\log\log T$. The expected reward from this is at least $(1 - \gamma_2(T, \epsilon))(1 - \epsilon)2\log\log T$. The optimal strategy has at least the expected reward of any strategy, therefore

$$\Psi(T) \geq (1 - \gamma_2(T, \epsilon))(1 - \epsilon)2\log\log T \tag{17}$$

for every $0 < \epsilon < 1/2$. $\square$

**Corollary 5.1.** *For $\log\log T \geq 4$ we have*

$$-32\log\log\log T - 8 \leq \Psi(T) - 2\log\log T \leq 8\log\log\log T + 6$$

The proof of the corollary is in the Appendix. The corollary implies that

$$\lim_{T \to \infty} \frac{\Psi(T)}{2\log\log T} = 1,$$

but in fact provides a much more refined convergence bound, showing that the low-order term is of the order of only $\log\log\log T$.

From Proposition 5 and (14), and observing that $\psi_t(\cdot)$ is minimized at zero, using Corollary 1, we derive the following upper bound on the threshold.

**Corollary 5.2.** *For $\log\log T \geq 35$,*

$$\theta(t) \leq t\sqrt{\psi_t(0)} \leq t\sqrt{2\log\log T + 8\log\log\log T + 6} \leq t\sqrt{3\log\log T}$$

However, the bound given above is far from tight. We conjecture that,

$$\theta(t) = \Theta(\sqrt{t\log\log t}). \tag{18}$$

Figure 1: Algorithm to compute optimal policy

**Algorithm 1.** *Parameters: Rectangle width $\gamma$, integration bounds $\ell, h$, maximum time $T$.*

1. *$\theta[t]$ is an array $[1..T]$, and $psi[t,i]$ is and array $[1..T, 1..T/\gamma]$*

2. *initialize: $\theta[1] \leftarrow 0$*

3. *functions: $x(i,c,t) := \sqrt{\frac{t-1}{t}}c - \sqrt{\frac{t}{t-1}}i\gamma$.*

   *$psi^*(t,i) := IF \ |i\gamma| < \theta[t] \ THEN \ return \ psi[t,i] \ ELSE \ return \ (i\gamma)^2/t$.*

4. *FOR $t = 2, 3, \ldots, T$*
   *j=0;*

   (a) *DO*
       *$c = j\gamma$; $StopValue \leftarrow c^2/t$*
       *$WaitValue \leftarrow \frac{1}{\sqrt{2\pi}} \sum\limits_{\ell/\gamma \le i \le h/\gamma} e^{-\frac{x_i^2}{2}} psi^*(t-1, i)\gamma$;*
       *$psi[t,j] \leftarrow \max\{WaitValue, StopValue\}$;*
       *UNTIL $StopValue \ge WaitValue$;*

   *$\theta(t) \leftarrow c$*
   *END-FOR*

---

# 6 Computing an Approximate Optimal Policy

## 6.1 Algorithm and Error Bound

We show how to approximate optimal policy by an efficient algorithm, based on the recurrence formula (6). Our algorithm (Figure 1) receives a parameter $\gamma > 0$ which controls its accuracy.

The algorithm iteratively uses the rectangle method, with given rectangle width $\gamma$, for approximating the integral in (8). Namely it approximates

$$\frac{1}{\sqrt{2\pi}} \int_{-\infty}^{\infty} e^{-\frac{x^2}{2}} \psi_{t-1}(b(x)) dx,$$

where $b(x) := \frac{t-1}{t}c - \sqrt{\frac{t-1}{t}}x$. Values of $\psi_{t-1}$ in this expression are taken from values computed in the previous iteration.

First, we bound the error in approximating the integral by truncating the tails. (See the Appendix for proof.)

**Lemma 6.1.** *For $\epsilon \in (0,1)$, $h \ge \sqrt{6 \log(2T/\epsilon)}$, and $\ell = -h$, we have*

$$\frac{1}{\sqrt{2\pi}} \int_{-\infty}^{\infty} e^{-\frac{x^2}{2}} \psi_{t-1}(b(x)) dx - \frac{1}{\sqrt{2\pi}} \int_{\ell}^{h} e^{-\frac{x^2}{2}} \psi_{t-1}(b(x)) dx \le \frac{\epsilon}{2T}$$

*where $b(x) := \frac{t-1}{t}c - \sqrt{\frac{t-1}{t}}x$*

Next we use a standard approximation using the rectangle method.

**Lemma 6.2.** *For $\gamma = \sqrt{\epsilon}/(T \log T \log \log T)$, and $h = -\ell = \sqrt{6 \log(2T/\epsilon)}$,*

$$\left| \frac{1}{\sqrt{2\pi}} \int_{\ell}^{h} e^{-\frac{x^2}{2}} \psi_{t-1}(b(x)) dx - \frac{1}{\sqrt{2\pi}} \sum_{\ell/\gamma < i < h/\gamma} e^{-\frac{x_i^2}{2}} \psi_{t-1}(i\gamma) \right| \le \frac{\epsilon}{2T}$$

where $x_i := \sqrt{\frac{t-1}{t}}c - \sqrt{\frac{t}{t-1}}i\gamma$.

*Proof.* We are using the rectangle method using width $\gamma$ over the interval $[\ell, h]$. The error for the rectangle method is bounded by $\frac{(h-\ell)\gamma^2}{24}K$ where $|\psi''_{t-1}(\xi)| \leq K$ for any $\xi \in [\ell, h]$. By Lemma D.1 we have that

$$K \leq \frac{3 + \theta^2(t)}{t} \leq 3t\log\log T + \frac{3}{t} \leq 4T\log\log T$$

This implies that the error is at most

$$\frac{1}{\sqrt{2\pi}}\frac{(h-\ell)\gamma^2}{24}K \leq \frac{1}{\sqrt{2\pi}}\frac{2\sqrt{6\log(2T/\epsilon)}}{24}\frac{\epsilon}{T^2\log^2 T(\log\log T)^2}4T\log\log T \leq \frac{\epsilon}{2T}$$

$\square$

**Proposition 6.1.** *Let $\gamma = \sqrt{\epsilon}/(T\log T\log\log T)$, and $h = -\ell = \sqrt{6\log(2T/\epsilon)}$. Then,*

$$\left|\frac{1}{\sqrt{2\pi}}\int_{-\infty}^{\infty}e^{-\frac{x^2}{2}}\psi_{t-1}(b(x))dx - \frac{1}{\sqrt{2\pi}}\sum_{\ell/\gamma<i<h/\gamma}e^{-\frac{x_i^2}{2}}\psi^*_{t-1}(i\gamma)\right| \leq \frac{\epsilon t}{T}$$

*where $b(x) := \frac{t-1}{t}c - \sqrt{\frac{t-1}{t}}x$ and $x_i := \sqrt{\frac{t-1}{t}}c - \sqrt{\frac{t}{t-1}}i\gamma$.*

*Proof.* By Lemma 6.1 and Lemma 6.2 we have,

$$\left|\frac{1}{\sqrt{2\pi}}\int_{-\infty}^{\infty}e^{-\frac{x^2}{2}}\psi_{t-1}(b(x))dx - \frac{1}{\sqrt{2\pi}}\sum_{\ell/\gamma<i<h/\gamma}e^{-\frac{x_i^2}{2}}\psi_{t-1}(i\gamma)\right| \leq \frac{\epsilon}{T}$$

We now show the proof by induction on $t$. Clearly it holds for $t = 1$. For the inductive step for $t-1$ we have $|\psi^*_{t-1}(i\gamma) - \psi_{t-1}(i\gamma)| \leq \epsilon(t-1)/T$. We now compute $\psi^*_t$ based on $\psi^*_{t-1}$ and have

$$\left|\psi^*_t(c) - \frac{1}{\sqrt{2\pi}}\sum_{\ell/\gamma<i<h/\gamma}e^{-\frac{x_i^2}{2}}\psi_{t-1}(i\gamma)\right| = \left|\frac{1}{\sqrt{2\pi}}\sum_{\ell/\gamma<i<h/\gamma}e^{-\frac{x_i^2}{2}}\psi^*_{t-1}(i\gamma) - \frac{1}{\sqrt{2\pi}}\sum_{\ell/\gamma<i<h/\gamma}e^{-\frac{x_i^2}{2}}\psi_{t-1}(i\gamma)\right|$$

$$\leq \frac{\epsilon}{T} + \frac{1}{\sqrt{2\pi}}\sum_{\ell/\gamma<i<h/\gamma}e^{-\frac{x_i^2}{2}}\left|\psi^*_{t-1}(i\gamma) - \psi_{t-1}(i\gamma)\right|$$

$$\leq \frac{\epsilon}{T} + \frac{\epsilon(t-1)}{T} = \frac{\epsilon t}{T}$$

where we used the fact that $\frac{1}{\sqrt{2\pi}}\sum_{\ell/\gamma<i<h/\gamma}e^{-\frac{x_i^2}{2}} \leq 1$.

$\square$

We now bound the running time of our algorithm.

**Proposition 6.2.** *Algorithm of Figure 1 runs in time $\tilde{O}(T^4/\epsilon)$ and computes an $\epsilon$ approximation of the optimal policy.*

*Proof.* Each computation of $WaitValue$ takes $(h-\ell)/\gamma = O(\sqrt{6\log(2T/\epsilon)}(T\log T\log\log T)/\sqrt{\epsilon}) = \tilde{O}(T/\sqrt{\epsilon})$ steps, and period $t$ calculates it $\theta(t)/\gamma = O(t\sqrt{\log\log T}(T\log T\log\log T)/\sqrt{\epsilon}) = \tilde{O}(T^2/\sqrt{\epsilon})$ times. There are $T$ periods, hence, the running time is $\tilde{O}((T/\sqrt{\epsilon})(T^2/\sqrt{\epsilon})T) = \tilde{O}(T^4/\epsilon)$. $\square$

As a supplementary step to the algorithm, the expert's expected optimal reward $\Psi(T)$ may be calculated from its definition in Definition 1, using numeric integration (e.g., with the rectangle method), without significant addition to error or running time.

Assuming that our conjectured bounds, as given in (18) and (25), hold, the algorithm is significantly faster: The integration rectangle width $\gamma$ can be widened by a factor of $O(\sqrt{T})$, for a gain of $O(T)$ in running time. Due to the lower bound on $\theta(t)$, the number of rectangles is reduced by a further factor of $O(\sqrt{T})$ for an overall $O(T^{1.5})$ improvement, and an algorithm completing in $\tilde{O}(T^{2.5}/\epsilon)$ steps

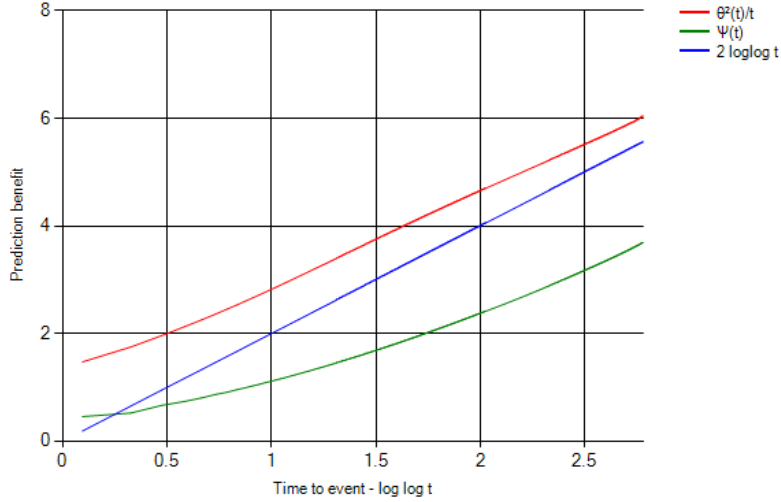

Figure 2: Optimal expert policy and expectation

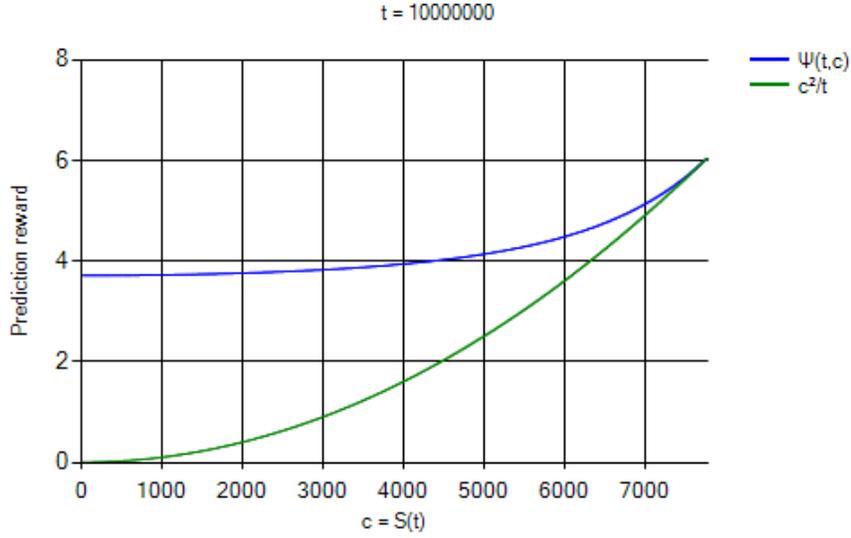

Figure 3: Canonical expectation dependency on signal

## 6.2 Empirical Simulation

We have derived the exact bound using an empirical method, and the results show that indeed our theoretical bounds are very close to the correct bounds.

The result of such a calculation, for up to $T = 10,000,000$ ($\ln \ln T \simeq 2.78$), is shown in Figure 2, where the expected reward is plotted against the time horizon (on a log-log scale). The figure has the optimal prediction threshold (shown in the form $\theta^2(t)/t$, i.e., the benefit of predicting with $|S_t| = \theta(t)$), the expert's expectation with $S_t = 0$, i.e., $\psi_t(0)$, and the reference $2 \log \log t$.

In Figure 3 the canonical expectation profile at $t = 10,000,000$ is shown for all values $S_t = c$ up to $c < \theta(t) \simeq 8,000$. For $t \geq \theta(t)$, the expectation equals $c^2/t$, which is plotted in reference.

# 7 Discussion

## 7.1 Conclusions

We analyzed the expert's policy in the prediction scenario described, and found that the expert's optimal strategy is to adopt a policy that aims to maximize the value of $(y_t - x_t)^2/t$. To do so the expert adopts a prediction threshold based on his deviation from the market value $|y_t - x_t| \geq \sqrt{q}\theta(t)$ where $q$ is his quality, and $\theta$ is a slowly increasing function of the time left $t$. His expected benefit from this optimal policy includes a positive term that depends on $q$ only, and a time-varying, slowly increasing term which is proportional to $q$ and a function $\psi_t(c)$ of the time left $t$ and his current deviation from the market $c$. The expert's expected optimal reawrd $\Psi(T)$ is roughly $2 \log \log T$, with bounds on the deviations.

Our results show that higher-quality experts should hold their silence where lower-quality experts are advised to speak. It also shows that an expert's reward expectation increases with the time left, albeit slowly. Receiving a large reward depends on having a prediction with a significant deviation from the market's, which may require waiting.

## 7.2 Structure of our random walk

Our random walk has an intricate structure which is neither a super-martingale nor a sub-martingale. Let us slightly elaborate more on the issue. As Section 3 shows, the expert aims to maximize $S_t^2/t$. A simple analysis (not shown) will show that when $S_t^2/t > 1$, $E[S_\tau^2/\tau|S_t] < S_t^2/t$ for all $\tau < t$, the definition of a super-martingale, while when $S_t^2/t < 1$, $E[S_\tau^2/\tau|S_t] > S_t^2/t$ for all $\tau < t$, the definition of a sub-martingale.

Does this not simply mean that the expert should use the threshold $S_t^2/t = 1$? After all, above this threshold, his target is expected to decrease, while below this threshold, his target is expected to increase. Waiting with $S_t^2/t$ above 1 would appear to have the same fault as the Martingale betting system, i.e., expecting to make gains in a losing situation.

The reason that this is not so is that the expert's future is not consistently a super-martingale, but will turn into a sub-martingale when crossing the 1 threshold. This virtually guarantees a reward of close to 1, a feature that is absent in the gambler's fallacy. When $S_t^2/t > 1$, the expert's downside is capped at close to 1, but his upside is unbounded. This motivates (up to a point, set by $\theta^2(t)/t$) the expert to wait in what is an apparently losing situation.

## 7.3 Other Random Walks and Scoring Rules

While our analysis used Gaussian random walks, and the logarithmic scoring rule, we believe, with some justification, that other random walks and other scoring rules will have similar qualitative results.

In particular, for other random walks, we note that Proposition 5 depends on the type of random walk only via Lemma 2.1. Once similar bounds are established for a random walk with non-Gaussian steps, the proof, with suitable modifications, may hold. For the Simple Symmetric Random Walk (with $\pm 1$ steps), we calculated an equivalent of the algorithm of Figure 1 in Section 6, to verify that its behavior differs only slightly from the behavior shown in Figures 2 and 3.

## 7.4 Future Work

An immediate open problem is a better characterization of the threshold function $\theta(t)$, hopefully showing that it is $\Theta(\sqrt{t \log \log t})$. However, there are many more conceptual open problems that we leave for future research.

While we gave a fairly precise characterization of the optimal expected reward of the expert, this does not characterize *when* the expert is likely to speak. More specifically: What is the mean time to prediction? With given initial deviation? Averaged over deviations? It appears that, on average, an expert will wait most of the time left. What if the expert is given a deadline for prediction? For example, if the expert must predict before $t = 10$, or before a given fraction of the initial time $t = T/k$?

Generalizing to more than one expert brings up a wide array of interesting questions, including how to model the correlation between the experts' signals. The expert's problem is greatly altered when all experts are strategic, and remains so even if we assume that only one expert is strategic, with the rest scheduled to disclose their predictions at predetermined times.

# A Proof of lemma 5.1

*Proof.* Define $\Phi(t) := \sqrt{2t \log \log T}$, $t \leq T$. Define $M_T := \max_{1 \leq t \leq T} S_t^2/t$. To establish an upper bound for $\Phi(T)$, we compute an upper bound for $\mathbb{E}[M_T]$, noting that $\mathbb{E}[M_T] \geq \Psi(T)$. Because, selecting (in hindsight) the $t$ that maximizes $S_t^2/t$ will always perform at least as well as any stopping strategy (which precludes hindsight).

For every $\epsilon > 0$ set $a := a(\epsilon) := 1 + \frac{(\sqrt{1+\epsilon} - \sqrt{1+\epsilon/2})^2}{1+\epsilon/2}$. Note that when $\epsilon > 0$, $a(\epsilon)$ is increasing and in the interval $(1, 4 - 2\sqrt{2})$.

By Lemma 2.1, for $\lambda = \sqrt{(1+\epsilon/2)2\log\log T}$, we have

$$\Pr[|S_t| \geq \Phi(t)\sqrt{1+\epsilon/2}] \leq (\log T)^{-(1+\epsilon/2)}$$

We bound $|S_t|$ for every $t = a^k \leq T$ with integer $k$ (in a slight abuse of notation, $a^k$ refers to its integer part). Using a union bound, we have,

$$\sum_{k=0}^{[\log_a T]} \Pr[|S_{a^k}| \geq \Phi(a^k)\sqrt{1+\epsilon/2}] < \frac{\log_a T}{(\log T)^{1+\epsilon/2}} = \frac{1}{(\log a)(\log T)^{\epsilon/2}} \tag{19}$$

Next we bound the deviations inside an interval. Define $\delta_k$ as follows

$$\begin{aligned}
\delta_k &:= \max_{t \in (a^k, a^{k+1}]} \Pr[|S_t - S_{a^k}| \geq \Phi(a^k)(\sqrt{1+\epsilon} - \sqrt{1+\epsilon/2})] \\
&\leq \max_{t \in (a^k, a^{k+1}]} \exp\left\{ -\frac{\Phi^2(a^k)(\sqrt{1+\epsilon} - \sqrt{1+\epsilon/2})^2}{2(t - a^k)} \right\} \\
&\leq \exp\left\{ -\frac{(\sqrt{1+\epsilon} - \sqrt{1+\epsilon/2})^2 2\log\log T}{2(a-1)} \right\} \\
&= \exp\left\{ -(1+\epsilon/2)\log\log T \right\} = (\log T)^{-(1+\epsilon/2)}
\end{aligned}$$

where the first inequality uses Lemma 2.1 over the $t - a^k$ time steps from $a^k$. The second inequality follows since $a^k/(t - a^k) > 1/(a-1)$ for $t \in (a^k, a^{k+1}]$.

By definition, $\delta_k$ bounds the probability that any single element will exceed the bound, but we need to bound the probability that the *maximum element* will exceed the bound. The following inequality comes very handy in relating the two quantities.

**Lemma A.1** (Levy's Inequality, see Lemma 1.2 in Damron (2012)). *Assume that for some $\delta > 0$ and $\lambda > 0$ we have $\Pr[|S_t| \geq \lambda/2] \leq \delta$ for all $t = 1, \ldots, T$. Then*

$$\Pr[\max_{t=1,\ldots,T} |S_t| \geq \lambda] \leq \frac{\delta}{1-\delta}$$

Since $S_t - S_{a^k}$ is distributed like $S_{t-a^k}$, using Lemma A.1 we have,

$$\Pr[\max_{t \in (a^k, a^{k+1}]} |S_t - S_{a^k}| \geq \Phi(a^k)(\sqrt{1+\epsilon} - \sqrt{1+\epsilon/2})] \leq \frac{\delta_k}{1-\delta_k} \leq \frac{2}{(\log T)^{1+\epsilon/2}},$$

where we use the fact that for $T > 10$ we have that $(\log T)^{-1} < 1/2$ and therefore $1 - \delta_k \geq 1/2$. Using union bound, the probability that for *any* $k$ we have $\max_{t \in (a^k, a^{k+1}]} |S_t - S_{a^k}| \geq \Phi(a^k)(\sqrt{1+\epsilon} - \sqrt{1+\epsilon/2})$, is at most

$$\frac{2\log_a T}{(\log T)^{1+\epsilon/2}} = \frac{2}{(\log a)(\log T)^{\epsilon/2}} \,. \tag{20}$$

Combining (19) and (20), assuming the high probability events hold, for $t \in (a^k, a^{k+1}]$ we have that,

$$|S_t| \leq |S_t - S_{a^k}| + |S_{a^k}| \leq \Phi(a^k)(\sqrt{1+\epsilon} - \sqrt{1+\epsilon/2}) + \Phi(a^k)\sqrt{1+\epsilon/2}$$
$$= \Phi(a^k)\sqrt{1+\epsilon} \leq \Phi(t)\sqrt{1+\epsilon}$$

holds for all $1 \leq t \leq T$ with probability at least $1 - \delta(\epsilon)$ where,

$$\delta(\epsilon) := \frac{1+2}{(\log a(\epsilon))(\log T)^{\epsilon/2}} = \frac{3}{(\log a(\epsilon))(\log T)^{\epsilon/2}}.$$

Rephrasing this in terms of $M_T$

$$\Pr[M_T > (1+\epsilon)2\log\log T] \leq \delta(\epsilon).$$

To bound $\mathbb{E}[M_T]$, by the definition of expectation

$$\mathbb{E}[M_T] - (1+\epsilon)2\log\log T \leq 2\log\log T \int_\epsilon^\infty \Pr[M_T > (1+x)2\log\log T]dx$$

Therefore

$$\mathbb{E}[M_T] - (1+\epsilon)2\log\log T \leq 2\log\log T \int_\epsilon^\infty \delta(x)dx$$

$$\leq \frac{6\log\log T}{\log a(\epsilon)} \int_\epsilon^\infty (\log T)^{-x/2}dx = \frac{6\log\log T}{\log a(\epsilon)} \int_\epsilon^\infty e^{-x(\log\log T)/2}dx$$

$$\leq \frac{6\log\log T}{\log a(\epsilon)} \left[ \frac{2}{\log\log T}(\log T)^{-x/2} \right]_\epsilon^\infty$$

$$= \frac{12}{(\log a(\epsilon))} \frac{1}{(\log T)^{\epsilon/2}} = \gamma_1(\epsilon)$$

This implies that $\Psi(T) \leq \mathbb{E}[M_T] \leq (1+\epsilon)2\log\log T + \gamma_1(\epsilon)$, which completes the proof of the lemma. $\qquad\square$

# B    proof of Lemma 5.2

*Proof.* Define $\Phi(t) := \sqrt{2t\log\log T}$, $t \leq T$. Define $M_T := \max_{1 \leq t \leq T} S_t^2/t$. By Lemma 2.1, for $\lambda = (1-\rho)\sqrt{2\log\log T}$, we have,

$$\Pr[S_t \geq (1-\rho)\sqrt{2t\log\log T}] \geq \frac{1}{(1-\rho)\sqrt{2\log\log T} + 2} \cdot \frac{1}{(\log T)^{(1-\rho)^2}} \tag{21}$$

As before, we consider the sequence of times $a^k$ with integer $k$. This time we set $a := a(\epsilon) := 20/\epsilon^2$. Assuming $\epsilon < 1/2$, this guarantees that,

$$(1+\epsilon/8)\Phi(a^{k-1}) \leq (\epsilon/4)\Phi(a^k) \tag{22}$$

and also

$$(1-\epsilon/8)\Phi(a^k - a^{k-1}) \geq (1-\epsilon/4)\Phi(a^k) . \tag{23}$$

We define $R_k := S_{a^k} - S_{a^{k-1}}$. Note that the $R_k$ are mutually independent and $R_k$ is distributed like $S_{a^k - a^{k-1}}$. Consider the events $\Pr[R_k \geq (1 - \epsilon/4)\Phi(a^k)]$ for every $k$ such that $a^k \leq T$. For each $R_k$ we have,

$$\Pr[R_k \geq (1 - \epsilon/4)\Phi(a^k)] = \Pr[S_{a^k - a^{k-1}} \geq (1 - \epsilon/4)\Phi(a^k)]$$
$$\geq \Pr[S_{a^k - a^{k-1}} \geq (1 - \epsilon/8)\Phi(a^k - a^{k-1})]$$
$$\geq \frac{(\log T)^{-(1-\epsilon/8)^2}}{(1 - \epsilon/8)\sqrt{2 \log \log T} + 2} \, ,$$

where the first inequality uses (23) and the second uses (21).

We need to lower bound the probability that at least one of these events, i.e., $R_k \geq (1 - \epsilon/4)\Phi(a^k)$, will occur. The probability that none will occur is at most

$$\left(1 - \frac{(\log T)^{-(1-\epsilon/8)^2}}{(1 - \epsilon/8)\sqrt{2 \log \log T} + 2}\right)^{\log_a T} < \delta_1 := \exp\left(-\frac{(\log T)^{\epsilon/4 - \epsilon^2/64}}{(1 - \epsilon/8)\sqrt{2 \log \log T} + 2} \cdot \frac{1}{\log a}\right)$$

since $(1 - b/x)^x < e^{-b}$ for $x > b > 0$.

Therefore with probability at least $1 - \delta_1$ for at least one $k$ the event occurs. We now show that, with high probability the value of $S_{a^k}$ is not too negative. By Lemma 2.1, for $\lambda = (1 + \epsilon/8)\sqrt{2 \log \log T}$, we have,

$$\Pr[S_{a^{k-1}} \leq -(1 + \epsilon/8)\Phi(a^{k-1})] \leq \delta_2 := (\log T)^{-(1+\epsilon/8)}$$

So the event $\Pr[\forall k : \ S_{a^{k-1}} > -(1+\epsilon/8)\Phi(a^{k-1})]$ occurs with probability at least $1 - \delta_3$ where $\delta_3 := (\log_a T)\delta_2$. Therefore, with probability at least $1 - \delta_1 - \delta_3$ we have

$$S_{a^k} = R_k + S_{a^{k-1}} \geq (1 - \epsilon/4)\Phi(a^k) - (1 + \epsilon/8)\Phi(a^{k-1})$$
$$\geq (1 - \epsilon/4)\Phi(a^k) - (\epsilon/4)\Phi(a^k)$$
$$= (1 - \epsilon/2)\Phi(a^k) \, ,$$

where the second inequality uses (22). Since $\Pr[M_T < 2(1 - \epsilon)\log \log T] \leq \Pr[M_T < 2(1 - \epsilon/2)^2 \log \log T]$, the lemma follows. $\qquad \square$

# C   proof of corollary 1

*Proof.* By Lemma 5.1 we have for any $\epsilon_1$,

$$\Psi(T) - 2\log \log T \leq 2\epsilon_1 \log \log T + \gamma_1(T, \epsilon_1).$$

We set $\epsilon_1 = \frac{4 \log \log \log T}{\log \log T}$. Using Eq. (16) we have that

$$\gamma_1(T, \epsilon_1) < \frac{96}{\epsilon_1^2} \frac{1}{(\log T)^{\epsilon_1/2}} = \frac{6(\log \log T)^2}{(\log \log T)^2} e^{-2 \log \log T} = \frac{6}{(\log \log T)^2}$$

This implies that

$$\Psi(T) - 2\log \log T \leq 8 \log \log \log T + \frac{6}{(\log \log T)^2} \leq 8 \log \log \log T + 6,$$

which, using the fact that $T \geq e^{16}$, gives the upper bound in the corollary.

By Lemma 5.3 we have for any $\epsilon_2$,

$$-(\gamma_2(T, \epsilon_2) + \epsilon_2)2 \log \log T \leq \Psi(T) - 2 \log \log T.$$

We set $\epsilon_2 = \frac{16 \log \log \log T}{\log \log T} < 1$.

$$\gamma_2(T, \epsilon) := \exp\Big(-\frac{(\log T)^{\epsilon/4 - \epsilon^2/64}}{(1 - \epsilon/8)\sqrt{2 \log \log T} + 2} \cdot \frac{1}{\log \frac{20}{\epsilon^2}}\Big) + \frac{1}{(\log T)^{\epsilon/8}}$$

$$\leq \exp\Big(-\frac{(\log T)^{\epsilon/8}}{\sqrt{2 \log \log T} + 2}\Big) + \frac{1}{(\log T)^{\epsilon/8}}$$

For $\epsilon_2$ we have

$$\gamma_2(T, \epsilon_2) \leq \exp\Big(-\frac{(\log \log T)^2}{\sqrt{2 \log \log T} + 2}\Big) + \frac{1}{(\log \log T)^2} \leq \frac{1}{\log T} + \frac{1}{(\log \log T)^2} \leq \frac{2}{(\log \log T)^2} \ ,$$

where we used the fact that $T \geq e^{16}$. This implies that

$$\Psi(T) - 2 \log \log T \geq -\Big(\frac{4}{(\log \log T)^2} + \frac{16 \log \log \log T}{\log \log T}\Big) 2 \log \log T = -32 \log \log \log T - 8 \ ,$$

again, using $T \geq e^{16}$. $\qquad\square$

# D  Second derivative

The following lemma bounds the second derivative of $\psi_t(c)$, a bound we will need in the Section 6.

**Lemma D.1.** *For every $t > 1$, $c$, $-\frac{\theta^2(t)}{t} < \psi_t''(c) < \frac{3 + \theta^2(t-1)}{t}$.*[4]

*Proof.* Refering to (6), whenever $\psi_t(c) = \frac{c^2}{t}$, $\psi_t''(c) = \frac{2}{t}$. This value is within the lemma bounds, proving the lemma for this case.

Otherwise, i.e., when $c < \theta(t)$, we differentiate (9) twice.

$$\psi_t''(c) = \frac{1}{\sqrt{2\pi \frac{t-1}{t}}} \int_{-\infty}^{\infty} \frac{d^2}{d^2 c}\Big[e^{-\frac{(x - \frac{t-1}{t}c)^2}{2\frac{t-1}{t}}}\Big] \psi_{t-1}(x) dx$$

$$= \frac{1}{\sqrt{2\pi \frac{t-1}{t}}} \int_{-\infty}^{\infty} \Big[(x - \frac{t-1}{t}c)^2 - \frac{t-1}{t}\Big] e^{-\frac{(x - \frac{t-1}{t}c)^2}{2\frac{t-1}{t}}} \psi_{t-1}(x) dx$$

$$= \frac{t-1}{t} \frac{1}{\sqrt{2\pi}} \int_{-\infty}^{\infty} (x^2 - 1) e^{-\frac{x^2}{2}} \psi_{t-1}\Big(\frac{t-1}{t}c - \sqrt{\frac{t-1}{t}}x\Big) dx$$

Therefore, by (8)

$$\frac{t}{t-1} \psi_t''(c) + \psi_t(c) = \frac{1}{\sqrt{2\pi}} \int_{-\infty}^{\infty} x^2 e^{-\frac{x^2}{2}} \psi_{t-1}\Big(\frac{t-1}{t}c - \sqrt{\frac{t-1}{t}}x\Big) dx \tag{24}$$

Since the right-hand side of (24) is positive, $\psi_t''(c) \geq -\frac{t-1}{t}\psi_t(c) \geq -\psi_t(c) \geq -\frac{\theta^2(t)}{t}$, as the lemma claims.

As for every $y$ $\psi_t(y) \leq \max(\frac{y^2}{t}, \frac{\theta^2(t)}{t}) \leq \frac{y^2}{t} + \frac{\theta^2(t)}{t}$, and $x^2 e^{-\frac{x^2}{2}}$ is positive, we infer from (24)

$$\frac{t}{t-1} \psi_t''(c) + \psi_t(c) \leq \frac{1}{(t-1)\sqrt{2\pi}} \int_{-\infty}^{\infty} x^2 e^{-\frac{x^2}{2}} \Big[\Big(\frac{t-1}{t}c - \sqrt{\frac{t-1}{t}}x\Big)^2 + \theta^2(t-1)\Big] dx$$

$$= \frac{1}{t-1}\Big[\Big(\frac{t-1}{t}c\Big)^2 + 3\frac{t-1}{t} + \theta^2(t-1)\Big]$$

---

[4] We can prove that $\psi_t''(c) \geq 0$, but skip this claim and its proof, as it contributes little to the issue at hand.

since the second, third and fourth moments of the standard normal distribution are $1, 0$ and $3$, respectively. As $\psi_t(c) > \frac{c^2}{t}$, the above leads to

$$\psi_t''(c) < \frac{3(t-1)}{t^2} + \frac{\theta^2(t-1)}{t} < \frac{3 + \theta^2(t-1)}{t}$$

as claimed. □

The bound given in Lemma D.1 seems to be far from tight. Though we cannot provide a proof, the following better bound seems to hold

$$\psi_t''(c) = O\Big(\frac{\log \log t}{t}\Big) \tag{25}$$

# E   Proof of Lemma 6.1

*Proof.* Note that,

$$\frac{1}{\sqrt{2\pi}} \int_{-\infty}^{\infty} e^{-\frac{x^2}{2}} \psi_{t-1}(b(x)) dx - \frac{1}{\sqrt{2\pi}} \int_{\ell}^{h} e^{-\frac{x^2}{2}} \psi_{t-1}(b(x)) dx =$$

$$\frac{1}{\sqrt{2\pi}} \int_{-\infty}^{\ell} e^{-\frac{x^2}{2}} \psi_{t-1}(b(x)) dx + \frac{1}{\sqrt{2\pi}} \int_{h}^{\infty} e^{-\frac{x^2}{2}} \psi_{t-1}(b(x)) dx,$$

and since $\ell = -h$ the value of the two summed integrals is identical. We consider the integral with $h$.

Recall the following simple identities:

$$\int_{h}^{\infty} e^{-z^2/2} dz = \Phi_c(h)\sqrt{2\pi} \tag{26}$$

$$\int_{h}^{\infty} z e^{-z^2/2} dz = [-e^{-z^2/2}]_h^{\infty} = e^{-h^2/2} \tag{27}$$

$$\int_{h}^{\infty} z^2 e^{-z^2/2} dz = [-z e^{-z^2/2}]_h^{\infty} + \int_{h}^{\infty} e^{-z^2/2} dz = h e^{-h^2/2} + \Phi_c(h)\sqrt{2\pi} \tag{28}$$

Therefore, for any quadratic function, by Lemma 2.1

$$\int_{h}^{\infty} (a_0 + a_1 z + a_2 z^2) e^{-\frac{z^2}{2}} dz = (a_0 + a_2)\Phi_c(h)\sqrt{2\pi} + (a_1 + a_2 h) e^{-h^2/2}$$

$$< \Big(\frac{\sqrt{2\pi}}{2}|a_0| + \frac{\sqrt{2\pi}}{2}|a_2| + |a_2|h + |a_1|\Big) e^{-\frac{h^2}{2}} \tag{29}$$

So, to bound $\int_{h}^{\infty} \psi_{t-1}(b(x)) e^{-\frac{x^2}{2}} dx$, we use the following inequality

$$\psi_t(b(x)) \leq \frac{b^2(x) + \theta^2(t)}{t} \leq \frac{c^2 + 2cx + x^2 + \theta^2(t)}{t}$$

Substituting in (29), with $|a_0| = \frac{c^2 + \theta^2(t-1)}{t-1}, |a_1| = \frac{2|c|}{t-1}, |a_2| = \frac{1}{t-1}$

$$\int_{h}^{\infty} \psi_{t-1}(b(x)) e^{-\frac{x^2}{2}} dx < \Big(\frac{\sqrt{2\pi} c^2 + \sqrt{2\pi}\theta^2(t-1)}{2(t-1)} + \frac{2|c|}{t-1} + \frac{3/2 + h}{t-1}\Big) e^{-\frac{h^2}{2}}$$

$$\leq \frac{\sqrt{2\pi}}{2(t-1)}\Big[c^2 + \theta^2(t-1) + \frac{4|c|}{\sqrt{2\pi}} + \frac{3}{\sqrt{2\pi}} + 2\sqrt{\frac{3}{\pi}\log\frac{2T}{\epsilon}}\Big]\Big(\frac{\epsilon}{2T}\Big)^3$$

Since $\epsilon < 1$, noting that $|c| \le \theta(t-1)$, and recalling that by Corollary 2 we have $\theta^2(t-1) < 3(t-1)^2 \log \log T$, then

$$\frac{1}{\sqrt{2\pi}} \int\limits_{h}^{\infty} \psi_{t-1}(b(x)) e^{-\frac{x^2}{2}} dx \le \frac{4\theta^2(t-1) + 2 + 2\epsilon^2 \sqrt{\log \frac{2T}{\epsilon}}}{(2(t-1))(2T^2)} \frac{\epsilon}{4T}$$

$$\le \left( \frac{15(t-1) \log \log T}{2T^2} + \frac{2 + 2\epsilon^2 \sqrt{\log \frac{2T}{\epsilon}}}{(2(t-1))(2T^2)} \right) \frac{\epsilon}{4T} \le \frac{\epsilon}{4T}$$

and the lemma follows for $T \ge 4$. $\qquad \square$